\documentclass[11pt,preprint]{aastex}

\def\dd{{\rm d}}

\def\be{\beta}
\def\Gn{\Gamma_n}
\def\bn{\beta_n}
\def\Gej{\Gamma_{\rm ej}}

\def\Gsh{\Gamma_{\rm sh}}
\def\lsh{l_{\rm sh}}
\def\bsh{\beta_{\rm sh}}
\def\Mej{M_{\rm ej}}

\def\Rdec{R_{\rm dec}}
\def\Rtr{R_{\rm trail}}

\def\mdec{m_{\rm dec}}
\def\be{\begin{equation}}
\def\ee{\end{equation}}

\def\dM{\dot{M}}
\def\mheat{m_{\rm heat}}
\def\Rb{R_\beta}
\def\mb{m_\beta}
\def\taub{\tau_\beta}

\def\tsw{t_{\rm sh}}
\def\En{E_{\rm n.f.}}
\def\Esh{E_{\rm sh}}
\def\Eh{E_{\rm heat}}
\def\bh{\beta_i}
\def\Erad{E_{\rm rad}}
\def\Eej{E_{\rm ej}}
\def\namb{n_{\rm amb}}
\def\Racc{R_{\rm acc}}
\def\tcool{t_{\rm cool}}
\def\Grel{\Gamma_{\rm rel}}
\def\tg{\tilde{\gamma}}

\newbox\grsign \setbox\grsign=\hbox{$>$} \newdimen\grdimen \grdimen=\ht\grsign
\newbox\simlessbox \newbox\simgreatbox \newbox\simpropbox
\setbox\simgreatbox=\hbox{\raise.5ex\hbox{$>$}\llap
     {\lower.5ex\hbox{$\sim$}}}\ht1=\grdimen\dp1=0pt
\setbox\simlessbox=\hbox{\raise.5ex\hbox{$<$}\llap
     {\lower.5ex\hbox{$\sim$}}}\ht2=\grdimen\dp2=0pt
\setbox\simpropbox=\hbox{\raise.5ex\hbox{$\propto$}\llap
     {\lower.5ex\hbox{$\sim$}}}\ht2=\grdimen\dp2=0pt

\def\simlt{\mathrel{\copy\simlessbox}}

\begin{document}

\title{Neutron-fed afterglows of gamma-ray bursts}

\author{Andrei M. Beloborodov\altaffilmark{1,2,3}} 

\altaffiltext{1}{Canadian Institute for Theoretical Astrophysics,
60 St. George Street, Toronto, ON M5S 3H8, Canada}

\altaffiltext{2}{Physics Department, Columbia University, 538  West 120th 
Street New York, NY 10027}

\altaffiltext{3}{Astro-Space Center of Lebedev Physical 
Institute, Profsojuznaja 84/32, Moscow 117810, Russia}

\begin{abstract}
Fireballs of gamma-ray bursts are partially made of free neutrons. 
Their presence crucially changes the mechanism of the fireball deceleration
by an external medium. At late stages of the explosion, neutrons
fully decouple from ions and coast with a constant Lorentz factor
$\Gamma_n$. As the ion fireball decelerates, the neutrons form a leading front.
This front gradually decays,
leaving behind a trail of decay products mixed with the ambient medium.
The kinetic energy of the decay products far exceeds the medium rest-mass 
energy, and the trail has a Lorentz factor $\gamma\gg 1$ at radii 
$R<\Rtr\approx 10\Rb\approx 10^{17}$~cm, where 
$\Rb\approx 10^{16}(\Gamma_n/300)$~cm is the mean radius of neutron decay.
The ion fireball sweeps up the trail material and drives a shock wave in it. 
Thus, observed afterglow emission is produced in the neutron trail. 
It can naturally re-brighten or display a spectral transition at 
$R\approx\Rtr$ where the impact of neutrons turns off.
Absence of any neutron signatures would point to an extremely low baryon 
loading of the fireballs and a strong dominance of a Poynting flux.
\end{abstract}



\section{Introduction}

The presence of neutrons in gamma-ray bursts (GRBs) and their 
possible role for the observed emission was realized in the past few years
(Derishev, Kocharovsky, \& Kocharovsky 1999a,b;
Bahcall \& M\'esz\'aros 2000; M\'esz\'aros \& Rees 2000; Fuller, Pruet,
\& Abazajian 2000; Pruet \& Dalal 2002). In an accompanying paper, 
we study in detail the nuclear composition of GRB fireballs and 
show that the presence of neutrons among the ejected baryons is 
practically inevitable (Beloborodov 2003, Paper~1). One implication is an 
observable multi-GeV neutrino emission from inelastic neutron-ion collisions 
in the fireball. Here, we focus on a different aspect. We show that the 
neutrons have a dramatic impact on the explosion dynamics at radii as large
as $10^{17}$~cm and propose a novel mechanism for GRB afterglow emission. 

Let us remind what happens in a relativistic explosion without 
neutrons (see M\'esz\'aros 2002 for a review). The ejected fireball with 
mass $\Mej$ and Lorentz factor $\Gej$ sweeps up an ambient medium and 
gradually dissipates its kinetic energy. The dissipation rate peaks at a 
characteristic 
``deceleration'' radius $\Rdec$ where half of the initial energy is dissipated.
This radius corresponds to swept-up mass $\mdec=\Mej/\Gej$. Further 
dynamics is described by the self-similar blast wave model of Blandford \& 
McKee (1976). How does this picture change in the presence of neutrons?


\section{Neutron front and its trail}  

Neutrons develop a Lorentz factor $\Gn=10^2-10^3$ at the very beginning 
of the explosion when the fireball is accelerated by radiation pressure 
(Derishev et al. 1999b). They are collisionally coupled to the ions in the 
early dense fireball, and decouple close to the end of the acceleration stage. 
Then the neutrons coast and gradually decay with a mean lifetime 
$\taub\approx 900$~s and a mean decay radius $\Rb=c\taub\Gn$,
\be
\label{eq:H}
 \Rb=0.8\times 10^{16}\left(\frac{\Gn}{300}\right) {\rm ~cm}.
\ee

At radii under consideration, $R>10^{15}$~cm, the ejected fireball is a 
shell of thickness $\Delta\ll R$. In contrast to neutrons, the ion component 
of the fireball is aware of the external medium and its Lorentz factor 
$\Gamma$ decreases. As $\Gamma$ decreases below $\Gamma_n$, the ions fall 
behind and separate from the neutrons. Thus the fireball splits into two 
relativistic shells which we name N (neutrons) and I (ions); the N-shell will 
also be called ``neutron front''. The I-shell lags behind by a distance $l$,
\be
\label{eq:l}
 \frac{l}{R}\approx \bn-\bh \approx \frac{1}{2\Gamma^2}-\frac{1}{2\Gn^2},
\ee
where $\bh$ and $\bn$ are the shell velocities in units of $c$.
For simplicity, we hereafter focus on the stage of complete separation 
$l>\Delta$ (it sets in right after the beginning of the I-shell deceleration 
if $\Delta\sim 10^{11}$~cm, and it covers the whole explosion if 
$\Delta\rightarrow 0$).

The mass of the N-shell gradually decreases because of the $\beta$-decay,
\be
\label{eq:Mn}
  M_n(R)=M_n^0\exp\left(-\frac{R}{\Rb}\right).
\ee
However, even at $R>\Rb$, the N-shell energy $E_n=\Gn M_nc^2$ is huge 
compared to the ambient rest mass $mc^2$. For example, at $R=\Rdec$, 
$E_n/\mdec c^2=X_n\Gn\Gej\exp(-\Rdec/\Rb)$ where $X_n=M_n^0/\Mej$.

The neutron decay products $p$ and $e^-$ share immediately their huge 
momentum with ambient particles due to two-stream instability
(the instability timescale is set by the ion plasma frequency
$\omega_p$ and it is the shortest timescale in the problem).
Thus, the  N-shell leaves behind a mixed trail with a relativistic 
bulk velocity $\beta<\bn$ which we calculate now.

Let $\dd m=(\dd m/\dd R)\dd R$ be ambient mass overtaken by the
N-shell as it passes $\dd R$ and $\dd M_n=(M_n/\Rb)\dd R$ be 
mass of decayed neutrons. The $\dd m$ and $\dd M_n$ share momentum and
form a trail element with proper mass $\dd m_*=\dd M_n+\dd m +\dd\mheat$
which includes heat dissipated in the inelastic $\dd m$-$\dd M_n$ collision.
The laws of energy and momentum conservation read
\begin{eqnarray}
\label{eq:laws}
   \Gn\dd M_n+\dd m=\gamma\dd m_*, \qquad
   \bn\Gn\dd M_n=\beta\gamma\dd m_*,
\label{eq:mom}
\end{eqnarray}
where $\gamma=(1-\beta^2)^{-1/2}$ is the trail Lorentz factor.
Let us denote
\be
\label{eq:zet}
  \zeta(R)=\frac{\dd M_n}{\dd m}
              =\frac{M_n}{\Rb}\left(\frac{\dd m}{\dd R}\right)^{-1}.
\ee
From equations~(\ref{eq:laws}) we find 
\be
\label{eq:beta}
  \beta(R)=\frac{\bn}{1+(\Gn\zeta)^{-1}},  \qquad
  \gamma(R)=\frac{\Gn\zeta+1}{(\zeta^2+2\Gn\zeta+1)^{1/2}}.
\ee 
It gives $\gamma\gg 1$ until essentially all neutrons have decayed. 
For illustration, let us specialize to a power-law 
density profile of the ambient medium. Then
\be
\label{eq:m}
  m(R)=\mb\left(\frac{R}{\Rb}\right)^k, \qquad
  \zeta(R)=\frac{M_n}{k\mb}\left(\frac{R}{\Rb}\right)^{1-k},
\ee
and $\zeta$ evolves from $\zeta=M_n^0e^{-1}/km_\beta\gg 1$ at $R=\Rb$ to 
$\zeta\ll 1$ at $R\gg\Rb$ as $M_n$ decays exponentially. There exists 
a characteristic radius $\Rtr$ where the trail becomes nonrelativistic
($\beta=0.5$). It is defined by condition $\zeta=\Gn^{-1}$ 
(see eq.~\ref{eq:beta}), which requires about 10 e-folds of the decay 
(for a typical $\mb\sim\mdec\sim 10^{-5} M_n^0$). Thus,
\be
 \Rtr\approx 10\Rb=0.8\times 10^{17}\left(\frac{\Gn}{300}\right){\rm ~cm}.
\ee
$\Rtr$ depends very weakly (logarithmically) on the ambient density and 
$X_n$.

Now that we know $\beta$, the trail density $n$ can be easily calculated. 
The neutron front has a small thickness $\Delta$ and the ambient particles 
cross it almost instantaneously (on timescale $\gamma^2\Delta/c\ll R/c$).
Measured in the N-shell frame, the flux of ambient particles is
$\bn\Gn n_0=(\bn-\beta)\Gn\gamma n_{\rm amb}$
where $n_0$ and $n_{\rm amb}$ are proper densities of the ambient particles 
ahead and behind the N-shell, respectively. This gives the medium 
compression as it is accelerated in the neutron front: 
$\namb/n_0=[\bn/\gamma(\bn-\beta)]$. The total density of the trail 
includes the neutron decay products, so $n=(1+\zeta)n_{\rm amb}$. 
Using equation~(\ref{eq:beta}), we get 
\be
\label{eq:n}
  \frac{n}{n_0}=\frac{\bn(1+\zeta)}{\gamma(\bn-\beta)}=
    (1+\zeta)\left(\zeta^2+2\Gn\zeta+1\right)^{1/2}.
\ee

The energy dissipated in the neutron front is given by
\begin{equation}
\label{eq:dEn}
  \dd\En=\gamma\left(\dd m_*-\dd m-\dd M_n\right)c^2
         =\left(\Gn-\gamma\right)\dd M_nc^2-(\gamma-1)\dd m\,c^2.
\end{equation}
After simple algebra (using eqs.~[\ref{eq:laws}]), we find
\be
\label{eq:En}
 \frac{\dd\En}{\dd R}=\left[\frac{\Gn\bn-\gamma\beta}
     {\Gn\gamma(\bn-\beta)}-1\right]\gamma\frac{\dd m}{\dd R}\, c^2.
\ee
It simplifies to $\dd\En=2\gamma^2\dd mc^2$ for the interesting regime
$1\ll\gamma\ll \Gn$ ($\Gn^{-1}<\zeta<\Gn$).

The trail is formed very hot. This can be seen when comparing 
inertial mass $\dd m_*$ (that includes heat) with rest mass 
$\dd M_b=\dd m+\dd M_n=(1+\zeta)\dd m$, which gives dimensionless 
relativistic enthalpy
\be 
\label{eq:mu}
  \mu=\frac{\dd m_*}{\dd M_b}=\frac{(\zeta^2+2\Gn\zeta+1)^{1/2}}{1+\zeta}.
\ee
For $\Gn^{-1}<\zeta<\Gn$ we find $\mu\gg 1$, i.e. internal energy 
of the trail far exceeds its rest-mass energy. 
The trail parameters are summarized in Table~1.

\begin{table}
\begin{center}
\caption[ ]{Trail parameters}
\end{center}
\begin{tabular}{ccccc} \hline\hline
&&& \\ [-3.5ex]
 & $\zeta>\Gn$ & $1<\zeta<\Gn$ & $\Gn^{-1}<\zeta<1$ & $\zeta<\Gn^{-1}$\\ [0.8ex]
\hline
&&& \\ [-3ex]
  $\gamma$  & $\Gn$ & $(\Gn\zeta/2)^{1/2}$ & $(\Gn\zeta/2)^{1/2}$ &
             $1$\\ [0.8ex]
  $n/n_0$   & $\zeta^2$ & $2\gamma\zeta$  &  $2\gamma$ &  $1$\\ [0.8ex]
  $\mu$     & $1$ & $\Gn/\gamma\approx 2\tg$  &  $2\gamma$ &  $1$\\ [0.8ex]
\hline
\end{tabular}
\end{table}

It is instructive to view the dissipation process
in the rest frame of the trail. Here,
the elements $\dd m$ and $\dd M_n$ have initial Lorentz factors 
$\gamma$ and $\tg=\Gn\gamma(1-\beta\bn)$, share their opposite momenta,
and come at rest. This is achieved via a plasma instability that 
isotropizes the particle momentum distribution. 
It may result in two isotropic ion populations in the trail: ambient ions 
with mean Lorentz factor $\gamma$ and $\beta$-decay protons with mean 
Lorentz factor $\tg$. 
Both populations are relativistically hot, $\gamma\gg 1$ and $\tg\gg 1$,  
as long as $\Gn^{-1}<\zeta<\Gn$. They have equal energies 
$\tg\dd M_n=\gamma\dd m$ (which correspond to equal relativistic bulk momenta 
of $\dd M_n$ and $\dd m$ before they come at rest in the trail frame).
The ratio of their temperatures is the reciprocal of their density ratio.

A detailed model of dissipation in the neutron front is an interesting 
plasma physics problem, which we defer to a future work. We emphasize here 
that it is different from dissipation in collisionless shocks.
The thickness of a shock, $\delta\sim c/\omega_p$, is set by the 
timescale of two-stream instability, and it is a discontinuity in the 
hydrodynamical sense. 
By contrast, the N-shell has thickness $\Delta\sim 10^{11}-10^{12}$~cm
which is orders of magnitude larger than $\delta$. The neutrons decay and 
cause dissipation everywhere in the N-shell. The ion medium velocity grows 
smoothly from 0 at the leading edge of the neutron front to $\beta$ at its 
trailing edge. This can be called ``volume dissipation''.


\section{Shock wave}

The ion fireball follows the neutron front and collects the trail.
As a result, (1) the ion Lorentz factor $\Gamma$ decreases and (2) a shock 
wave propagates in the trail material. The shock has a Lorentz factor 
$\Gamma\simlt\Gsh<\Gn$ and cannot catch up with the neutron front 
(unless $n_0[R]$ falls off steeper than $R^{-3}$). The shock lags behind
the neutron front by distance $\lsh\approx R(\bn-\bsh)$ and its time lag is
\be
\label{eq:tsw}
  \tsw=\frac{\lsh}{c(\bsh-\beta)}\approx\frac{\gamma^2}{(\Gsh^2-\gamma^2)}
        \frac{R}{c}.
\ee
It is much shorter than $R/c$ as long as $\Gsh\gg\gamma$. Hereafter we
use the approximation $\tsw<R/c$ and assume that the trail is quickly picked 
up by the shock, before it could expand and change its density or velocity.
We allow, however, for rapid radiative losses of the trail: it 
may cool on a timescale $\tcool\ll R/c$ and possibly $\tcool<\tsw$. 
Below we consider two extreme regimes: $\tcool\gg\tsw$ 
(adiabatic) and $\tcool\ll\tsw$ (radiative with $\mu=1$).

In the radiative regime, the energy dissipated by the shock is
\be
  \frac{\dd\Esh}{\dd R}=\Gamma(\Grel-1)(1+\zeta)\frac{\dd m}{\dd R}\,c^2,
\ee
where $\Grel=\Gamma\gamma(1-\bh\beta)$ is the trail Lorentz factor with 
respect to the I-shell. For $\Gn^{-1}\ll\zeta\ll\Gn$ one can use 
approximate expressions $\zeta=2\gamma^2/\Gn$ and 
$\Grel=(1/2)(\gamma/\Gamma+\Gamma/\gamma)$ to get
$\dd\Esh=(1/2\gamma+\gamma/\Gn)(\Gamma-\gamma)^2 \dd m\, c^2$.
Note that the dissipation is smaller than it would be in the absence of 
the neutron front, $\dd\Esh=\Gamma(\Gamma-1)\dd m\, c^2$. Hence, in the 
radiative regime, the neutrons delay the deceleration of the ion fireball. 

In the adiabatic regime, the preshock material has a high enthalpy $\mu$ 
given by equation~(\ref{eq:mu}). The postshock heat, measured in the lab 
frame, is $\dd\Eh=\Gamma(\Grel\mu-1)(1+\zeta)\dd m$. It includes heat 
deposited into the preshock medium by the neutron front (eq.~\ref{eq:dEn}), 
which should be subtracted to get the energy dissipated in the shock itself.
Hence,
\be
\label{eq:Esh}
  \frac{\dd\Esh}{\dd R}=\left[\Gamma(\Grel\mu-1)-\gamma(\mu-1)\right]
   (1+\zeta)\frac{\dd m}{\dd R}\,c^2.
\ee
For $\Gn^{-1}\ll\zeta\ll\Gn$ we get $\dd\En=2\gamma^2\dd m c^2$,
$\dd\Esh=(\Gamma^2-\gamma^2)\dd m c^2$, and the total dissipated energy 
$\dd\Eh=(\Gamma^2+\gamma^2)\dd m c^2$. As long as $\Gamma\gg \gamma$, the 
bulk of energy is dissipated in the shock rather than in the neutron front. 
Moreover, the dissipated energy is the same as in the absence of a neutron 
front, and hence the fireball deceleration is the same. 

In both radiative and adiabatic regimes, the shock dissipation is 
suppressed when the blast wave decelerates to $\Gamma\sim\gamma$.
At $\Gamma=\gamma$ the shock would disappear.
Thus, $\Gamma$ is bound from below by $\gamma$.


\section{Numerical examples}

Let us consider the simple radiative regime. The I-shell and the shocked 
part of the trail can be treated as a single shell with a growing mass 
$M(R)=\Mej+m-M_n$ (a known function of radius) and a decreasing $\Gamma$.
Mass gain $\dd M$ causes deceleration $\dd\Gamma$ that is found from the 
energy and momentum conservation: $\dd(\Gamma M)=\gamma\dd M -\dd\Erad/c^2$ 
and $\dd(\bh\Gamma M)=\beta\gamma\dd M-(\dd\Erad/c^2)\bh$ where 
$\dd\Erad=\dd\Esh$ is the radiated energy. Excluding $\dd\Erad$,
we get the dynamic equation, 
\be
\label{eq:dyn}
 M\frac{\dd\Gamma}{\dd R}
   =-\Gamma^2\gamma\bh(\bh-\beta)(1+\zeta)\frac{\dd m}{\dd R}.
\ee
Using equations~(\ref{eq:beta}) and (\ref{eq:m}) we can solve numerically
equation~(\ref{eq:dyn}) for $\Gamma(R)$ with an initial condition $\Gej$. 
The results depend on the type of the ambient medium. 
Figure~1 shows two examples: a standard ISM and a wind from a Wolf-Rayet 
progenitor.

\begin{figure}
\begin{center}
\plotone{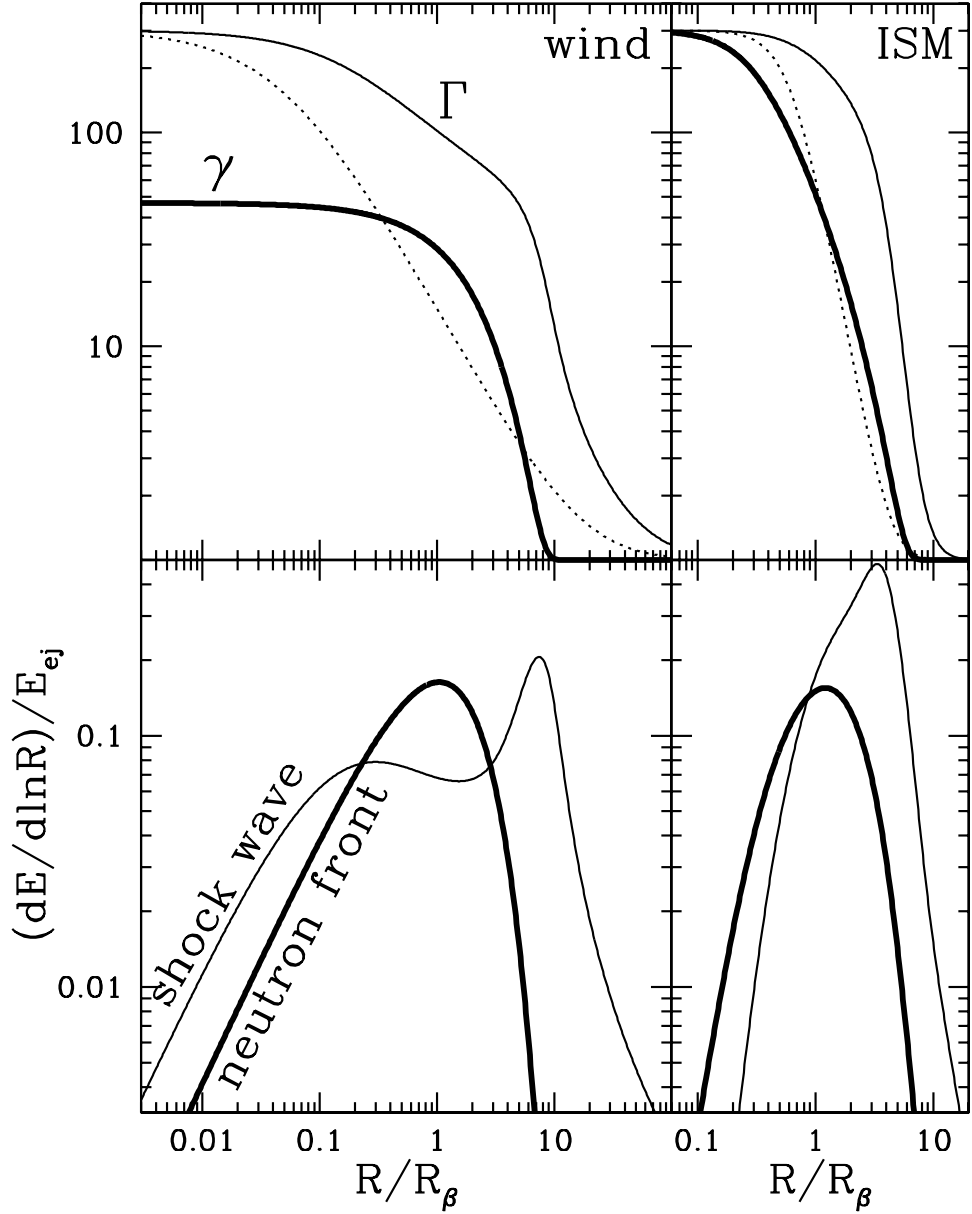}
\end{center}
\caption{Radiative model with $X_n=0.5$ and $\Gn=\Gej=300$. The ambient
medium is parametrized by $\mb$ and $k$ (eq.~\ref{eq:m}). Left panels show
the case of $k=1$ and $\mb=10\Mej/\Gej$ (typical for a wind from a
Wolf-Rayet progenitor). Right panels show the case of a normal interstellar
medium with $k=3$ and $\mb=\Mej/\Gej$. {\it Top:} Lorentz factors of the
neutron trail ($\gamma$) and ion fireball ($\Gamma$). Dotted curve shows the
fireball deceleration $\Gamma(R)$ that would take place without neutrons.
{\it Bottom:} Radial distributions of dissipation rates in the neutron front
and the shock.  $\Eej=\Gej\Mej c^2$ is an initial total energy of the ejecta.
$R$ is measured in units of the mean decay radius (eq.~1).
}
\end{figure}

The radiative explosion has two separate emission regions: behind the 
neutron front and behind the shock front. Their luminosities equal the 
corresponding dissipation rates $(c\,\dd\En/\dd R)$ and $(c\,\dd\Esh/\dd R)$ 
(Fig.~1). The neutron front dissipation peaks at $\Rb$. The shock dissipation 
can have two peaks (if $\mb\gg\Mej/\Gej$ as in the wind example in Fig.~1). 
The first peak  marks the beginning of the I-shell deceleration, and it is 
followed by a minimum when $\Gamma$ approaches $\gamma$.  At $R>2\Rb$, 
$\gamma$ falls down steeply and the shock becomes strong again (high 
$\Grel$), leading to the second peak at $R\approx\Rtr$.


\section{Discussion}

Dynamics of neutron-fed explosions is a clean physical problem: 
existence of neutrons in GRBs and their mean lifetime of 15~min 
is all we needed to construct the model. We focused here on the simple case 
where all neutrons have equal Lorentz factor $\Gn$, and data may require 
a multi-shell picture with variable $\Gn$. This extension could affect our 
model in case of high variations $\Delta\Gn>\Gn$ (though at large $R$, where
neutrons die out exponentially, it is sufficient to consider a highest 
$\Gn^{\rm max}$ as slower neutrons have shorter lifetimes and their 
population is smaller by $\exp[-\Gn^{\rm max}/\Gn]$). Another simple 
extension is beamed ejecta. Then at late stages the ion fireball spreads 
laterally while the neutron beaming remains constant. 

In contrast to dynamics, emission is really 
complicated. In general, radiation from collisionless blast waves is not 
derived from first principles, and neutrons do not make the problem simpler.
The afterglow emission is believed to be synchrotron, and it depends on 
poorly understood generation of magnetic field and electron acceleration.
The standard model without neutrons relies on the field generation in the 
shock by the two-stream instability (Sagdeev 1966, Medvedev \& Loeb 1999, 
Gruzinov 2001). The shock front is, however, extremely thin 
($\delta\sim c/\omega_p\ll R/\Gamma^2$), and the postshock field decays 
quickly. The model needs a significant remnant field in an extended layer 
behind the shock, which is uncertain. This problem is alleviated in 
neutron-fed explosions. Here, the leading neutron front is an additional 
dissipation region maintained in a turbulent state with generated magnetic 
fields. The N-shell has thickness $\Delta$ comparable to $R/\Gn^2$.
It can produce a significant synchrotron radiation by 
itself\footnote{The N-shell emission decays exponentially on observed 
timescale $\Rb/c\Gn^2=\taub/\Gn\sim 1-10$~s and it can be related to smooth
(fast-rise-exponential-decay) pulses in prompt GRBs.}
and leave behind remnant fields and hot plasma to be used by the ensuing 
shock wave for afterglow emission. We emphasize differences from a 
customary external shock: the neutron-trail shock propagates in a 
relativistically moving, dense, hot, and possibly magnetized medium. 
The fate of magnetic fields in such shocks and the resulting emission is 
an interesting issue for a future study.

The neutron impact ceases at $\Rtr\approx 10^{17}$~cm, which can leave an 
imprint on the observed afterglow. For example, the shock dissipation can 
have a bump (Fig.~1), and a spectral transition is also possible. 
The arrival time of radiation emitted at $\Rtr$ is $\approx\Rtr/2\Gamma^2c$ 
(counted from the arrival of first $\gamma$-rays). It may be as long as 30 
days or as short as a few seconds, depending on the fireball Lorentz factor 
$\Gamma(\Rtr)$. Remarkably, $\Rtr$ is almost independent of the ambient 
medium, and its observational signature would give information on 
$\Gamma(\Rtr)$. Recent early 
observation of a GRB afterglow (GRB~021004) discovered an interesting 
re-brightening at $10^3$~s which would correspond to 
$\Gamma(\Rtr)\approx 30(\Gn/300)^{1/2}$.
Also, we do not exclude a possible relevance of neutrons to the 20~day 
bumps observed in a few GRBs, as the time coincides with $\Rtr/c$. 

Neutron signatures should be absent if the fireball
is strongly dominated by a Poynting flux and has extremely low baryon 
loading. Then the neutron component decouples early, with a modest 
Lorentz factor $\Gn$, and decays at small radii. The upper bound on $\Gn$ 
due to decoupling is $\Gn\approx 300(\dM_\Omega/10^{26})^{1/3}$ where 
$\dM_\Omega$ [g/s] is the mass outflow rate per unit solid angle of the 
fireball (see \S~4.1 in Paper~1).

We focused here on the neutron front and did not account for the 
$\gamma$-ray precursor that impacts the blast wave dynamics at 
$R<\Racc=0.7\times 10^{16}(E_\gamma/10^{53})^{1/2}$~cm, where $E_\gamma$ 
[erg] is the isotropic energy of the GRB (see Beloborodov 2002 and 
refs. therein). The analysis in this Letter is strictly valid for 
afterglows emitted at $R>\Racc$. Then the radiation-front effects, 
including the gap opening at $R<0.3\Racc$, occur at smaller radii, and 
apply to an earlier afterglow. For a dense medium, where $\Rdec<\Racc$, 
effects of the neutron and $\gamma$-ray fronts should be studied together.

\acknowledgments
I am grateful to P.~Goldreich, R.~Blandford, and TAPIR group for their 
hospitality during my stay at Caltech, where part of this work was done.
This research was supported by NSERC and RFBR grant 00-02-16135.


\end{document}